\documentclass[journal]{IEEEtran}
\usepackage{amssymb}
\usepackage{amsmath}      
\usepackage{cite}         
\usepackage{algorithmicx} 
\usepackage{graphicx}     
\usepackage{tabularx, booktabs} 
\usepackage{subfig}
\usepackage{caption}
\usepackage{xcolor}

\begin{document}

\title{Fusion of Microgrid Control with Model-free Reinforcement Learning: Review and Vision}

\author{Buxin She,~\IEEEmembership{Student Member,~IEEE,}
        Fangxing Li,~\IEEEmembership{Fellow,~IEEE,}
        Hantao Cui,~\IEEEmembership{Senior Member,~IEEE,}
        \\ Jingqiu Zhang, ~\IEEEmembership{Student Member,~IEEE,}
        Rui Bo, ~\IEEEmembership{Senior Member,~IEEE}
                \vspace{-1em}
}

\markboth{IEEE Transaction on Smart Grid}%
{Shell \MakeLowercase{\textit{et al.}}: Bare Demo of IEEEtran.cls for IEEE Journals}

\maketitle

\begin{abstract}
Challenges and opportunities coexist in microgrids as a result of emerging large-scale distributed energy resources (DERs) and advanced control techniques. In this paper, a comprehensive review of microgrid control is presented with its fusion of model-free reinforcement learning (MFRL). A high-level research map of microgrid control is developed from six distinct perspectives, followed by bottom-level modularized control blocks illustrating the configurations of grid-following (GFL) and grid-forming (GFM) inverters. Then, mainstream MFRL algorithms are introduced with an explanation of how MFRL can be integrated into the existing control framework. Next, the application guideline of MFRL is summarized with a discussion of three fusing approaches, i.e., model identification and parameter tuning, supplementary signal generation, and controller substitution, with the existing control framework. Finally, the fundamental challenges associated with adopting MFRL in microgrid control and corresponding insights for addressing these concerns are fully discussed.
\end{abstract}

\section{Introduction}

\IEEEPARstart{M}{icrogrids} are gaining popularity due to their capability for accommodating distributed energy resources (DERs) and form a self-sufficient system  \cite{olivares2014trends}. Microgrids not only contribute to the development of a zero-carbon city but also work as a fundamental component of the ‘source, network, and load’ integrated energy systems. A microgrid may incorporate various types of energy sources and act as an energy router \cite{farrokhabadi2019microgrid}, making it possible for the grid to survive severe events while also making the country more energy-resilient and secure \cite{liu2016enhanced}.

A typical microgrid is composed of various DERs, energy storage systems, and loads that are connected locally as a united controlled entity\cite{chen2021peer}. In comparison to a traditional synchronous generator-dominated bulk power system, microgrids have a larger penetration of DERs \cite{li2012autonomous}-\cite{she2022virtual}, a smaller system size \cite{ju2017two}, a greater degree of uncertainty \cite{rezkallah2017microgrid}, lower system inertia \cite{nikmehr2015optimal} - \cite{she2022time}, and a stronger coupling of voltage and frequency (V-f). All these features pose challenges to the design of a microgrid control system. A complete microgrid control system is comprised of software and hardware that can both perform microgrid functionalities and guarantee stability at the same time  \cite{bidram2012hierarchical}. The software is also referred to as microgrid controllers, and focuses on control algorithm design in the paper. Existing microgrid controllers are usually designed under a hierarchal framework that includes the primary, secondary, and tertiary controllers\cite{wu2014control}. Ref. \cite{adineh2020review} conducted a thorough review of the hierarchal control of microgrids. There are also some articles providing an overview from the different perspectives of communication interfaces \cite{guerrero2012advanced}, operation modes \cite{andishgar2017overview}, and control techniques \cite{ahmed2020stability}. All these reviews provided an excellent summary and future directions of microgrid control research. As a result, we synthesize the valuable viewpoints and develops a high-level research map of microgrid control based on existing work. Furthermore, modularized control blocks have been developed to dive into the design of the fundamental units of microgrids: grid-following (GFL) and grid-forming (GFM) inverters\cite{han2017mas}, which is advantageous for microgrid researchers.

Model-free controllers have been used previously in microgrid control because they are easy to set up and independent of the physical model of the microgrid components. For example, fuzzy logic controllers \cite{deshmukh2020fuzzy} -\cite{garcia2018decentralized} and adaptive controllers \cite{zhang2015data} -\cite{zhang2021prescribed} can adjust their output based on pre-defined membership functions and adaption laws, respectively. However, they are difficult to scale up and cannot deal with emerging uncertainties in microgrids. Neural network control \cite{rodriguez2020very} -\cite{lin2021voltage} is another type of well-known model-free method. Although neural network is good at perception and decision-making based on historical data, it lacks exploration capability and cannot adapt to the rapidly changing microgrid environment. Apart from the above-mentioned model-free techniques, reinforcement learning (RL) is a prominent approach that is concerned with how an intelligent agent learns to solve Markov Decision Processes (MDP) in an environment. If we do not assume knowledge or an exact mathematical model of the environment, RL is referred to as model-free reinforcement learning (MFRL). Then, the RL agent finds the optimal policy through repeated interactions with the environment \cite{du2021intelligent}-\cite{du2022demonstration}. MFRL is a promising data-driven and model-free approach since it is not dependent on an accurate system model and does not need as many labeled datasets as supervised learning. In addition, it has strong exploration capability and can achieve autonomous operation once set up. MFRL is gaining more and more attention due to its successful applications in video games \cite{peng2020understanding}, autonomous driving \cite{kiran2021deep}, robotics \cite{brunke2021safe}, and power systems \cite{zhang2019deep}. Recently, researchers from DeepMind and École Polytechnique Fédérale de Lausanne developed a non-linear, high-dimensional, and RL-based magnetic controller for nuclear fusion \cite{degrave2022magnetic} and published their work in $Nature$. This indicates the great potential of implementing MFRL in engineering control (microgrid control).

For now, MFRL is still under development and needs further study. While some research has been conducted on MFRL for its application in microgrid control, there has been no in-depth review of how MFRL can be integrated into the current microgrid control framework. Hence, this paper performs a comprehensive review of the control framework of microgrids and summarizes how MFRL fuses with the existing control schemes.

Compared with other review papers on microgrid control, the main merits of this manuscript include:

•	Plotting of a high-level research map of microgrid control from the perspective of operation mode, function grouping, timescale, hierarchical structure, communication interface, and control techniques.

•	Development of modularized control blocks to dive into the fundamental units of microgrids: GFL and GFM inverters.

•   Introduction of the mainstream MFRL algorithms and summary of MFRL application guidelines, and the answering of two important questions: \textit{i).‘What kinds of tasks is MFRL suitable for?’}; \textit{ii).‘How can MFRL be fused with the existing microgrid control framework?’}.

•	Discussion of the primary challenges associated with adopting MFRL in microgrid control and providing insights for addressing these concerns.

The rest of this paper is organized as follows. Section II introduces the current microgrid control framework, including a high-level research map and modularized control blocks. Section III gives a brief introduction to RL and the mainstream algorithms of MFRL. The characteristics of each algorithm and its application scenarios in microgrid control are also summarized. A full discussion of the fusion of microgrid control with MFRL is presented in Section IV, along with the associated challenges and insights. Section V concludes this paper.

\section{Microgrid control framework}
This section first plots a high-level research map of microgrid control, and then develops modularized control blocks to dive into GFL and GFM inverters.

\subsection{High-level research map of microgrid control}

\begin{figure*}[htbp]
 \centering
   \vspace{-0.35cm}
   \includegraphics[scale=0.8]{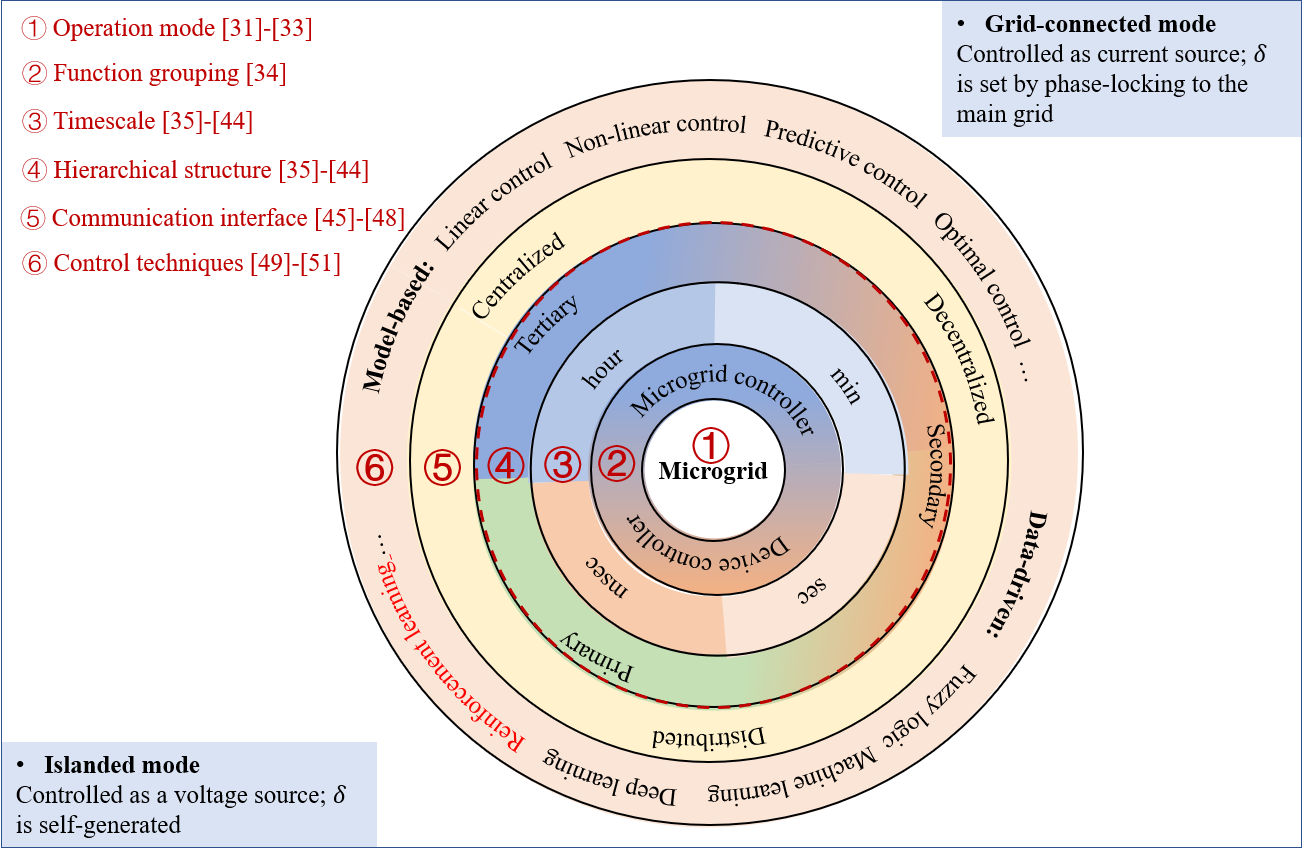}
   \vspace{-0.15cm}
 \caption{High-level research map of microgrid control}
 \label{HighLevelMap}
\end{figure*}

 Fig. \ref{HighLevelMap} shows the high-level research map of microgrid control from the perspectives of 1) operation mode, 2) function grouping, 3) timescale, 4) hierarchical structure, 5) communication interface, and 6) control techniques. For each perspective, there are articles providing comprehensive reviews. They are denoted in Fig. \ref{HighLevelMap} for the reader’s reference.

\subsubsection{Operation mode}
A microgrid can operate in either grid-connected (GC) mode or islanded (IS) mode depending on its connectivity to the main grid \cite{anderson2019review} -\cite{hathiyaldeniye2022optimal}. In GC mode, the microgrid keeps tracking the phase of the main grid through the phase-locking loop (PLL), and exchanges the mismatched power at the point of common coupling (PCC). In IS mode, the microgrid forms a self-sufficient system based on the local generations. Ref. \cite{d2020microgrid} summarized the strategies for the seamless transition between GC and IS modes.

\subsubsection{Function grouping}
To meet the objectives of the microgrid operation, the $2^{nd}$ viewpoint is associated with function grouping, which specifically include the microgrid controller and device controller \cite{ieee2017ieee}. 
Grid-level controllers focus on supervisory control functions and grid interactive control functions, and they are more likely to be software-based and applied to the hardware; while device controllers focus on device-level control functions and local-area control functions, and they are more likely to be applied directly on the hardware (devices and assets).

\subsubsection{Timescale}
The time scale of microgrid control is tightly related with the control structure. So, it will be discussed in detail in the next discussion about hierarchical structure.

\subsubsection{Hierarchical structure}
The hierarchical control structure is another specific function grouping perspective that clearly sets up the control targets for all the controllers, with which each level controller can work independently within the distinct timescales \cite{bidram2012hierarchical}.

The primary controller is responsible for voltage and current control of inverters and automatic power sharing among generations while maintaining V-f stability on a timescale of seconds \cite{she2022decentralized}. The indirect current control is used in the early stages \cite{du2019comparative}-\cite{li2010adaptive}, and is later replaced by the direct current control due to its fast response and accurate current control capability \cite{rocabert2012control}. More details can be found in the review paper \cite{rokrok2018review}. Because the primary controller pertains to fast control actions, it predominantly determines the stability of microgrids \cite{farrokhabadi2019microgrid}. Ref. \cite{khayat2019secondary} gave an overview of the primary control of microgrids. The secondary controller mitigates the V-f deviation unsolved by the primary controller in the timescale of seconds to minutes. It improves the power quality by generating supplementary signals based on the errors between the measurements and reference values. Ref. \cite{singh2019review} -\cite{xie2021optimization} performed a review on the secondary control of ac microgrids. The tertiary controller mainly focuses on economic and resilient operations in the timescale of minutes to hours. It adjusts the setting points of the primary and secondary controllers by solving optimal power flow and considering the load side demand response. Some reviews can be found in \cite{kanakadhurga2022demand} -\cite{almada2016centralized}.

\subsubsection{Communication interface}	
Depending on the communication interface, the control structure of the microgrid can also be categorized into centralized control, decentralized control, and distributed control\cite{espina2020distributed}. 

In centralized control, the microgrid control center coordinates the load and generation and responds to all disturbances. It collects and processes all the local information before sending the control signals to each device. The centralized control has the advantage of accurate power-sharing and good transient performance but suffers from the high cost of the communication device and single point failure. In distributed control, each node communicates only with its adjunct nodes. Average-based, consensus-based, and event-triggered distributed algorithms are employed in microgrid control \cite{singhal2022consensus}. Distributed control algorithms require the connected communication graph of microgrids. They also have a reduced convergence speed as the network grows \cite{sahoo2020communication}. In decentralized control, the control signals are generated based on local measurements. It has the advantage of the plug-and-play capability and is free of communication channel time delay, but it suffers from inaccurate power-sharing and large V-f deviation after disturbances. Ref. \cite{babayomi2022advances} conducted a review from the perspective of communication interfaces and summarized some tricks to address their flaws.

\subsubsection{Control techniques}	
Both model-based and data-driven control techniques have been utilized in microgrid control. Beginning with the classical linear control theory, advanced model-based control approaches such as non-linear control, optimum control, and model-predictive control (MPC) are then extensively used in microgrids. Ref. \cite{mohammadi2021robust} summarized the advances and opportunities of employing MPC in microgrids, and \cite{moharm2019state} reviewed the robust control strategies in microgrids. To address the problems of model uncertainty and unavailability, a variety of data-driven methodologies such as cutting-edge machine learning (ML) and deep learning (DL) are also employed in microgrid control. Ref. \cite{aslam2021survey} reviewed the application of big data in microgrids, and \cite{du2019comparative} conducted a survey on DL for microgrid load and DER foresting. A review of MFRL for microgrid control has yet to be done, which is why it is the main scope of this manuscript. 

In summary, MFRL is a promising approach that is worth investigating and being employed in microgirds. As shown in the high-level research map, MFRL doesn't mean to replace the existing control framework, but to complement it, improve it in a data-driven way, and finally work as an integrated part of the microgrid controller.

\subsection{Configuration of grid-following and grid-forming inverters}
\begin{figure*}[htbp]
 \centering
   \vspace{-0.25cm}
   \includegraphics[scale=0.33]{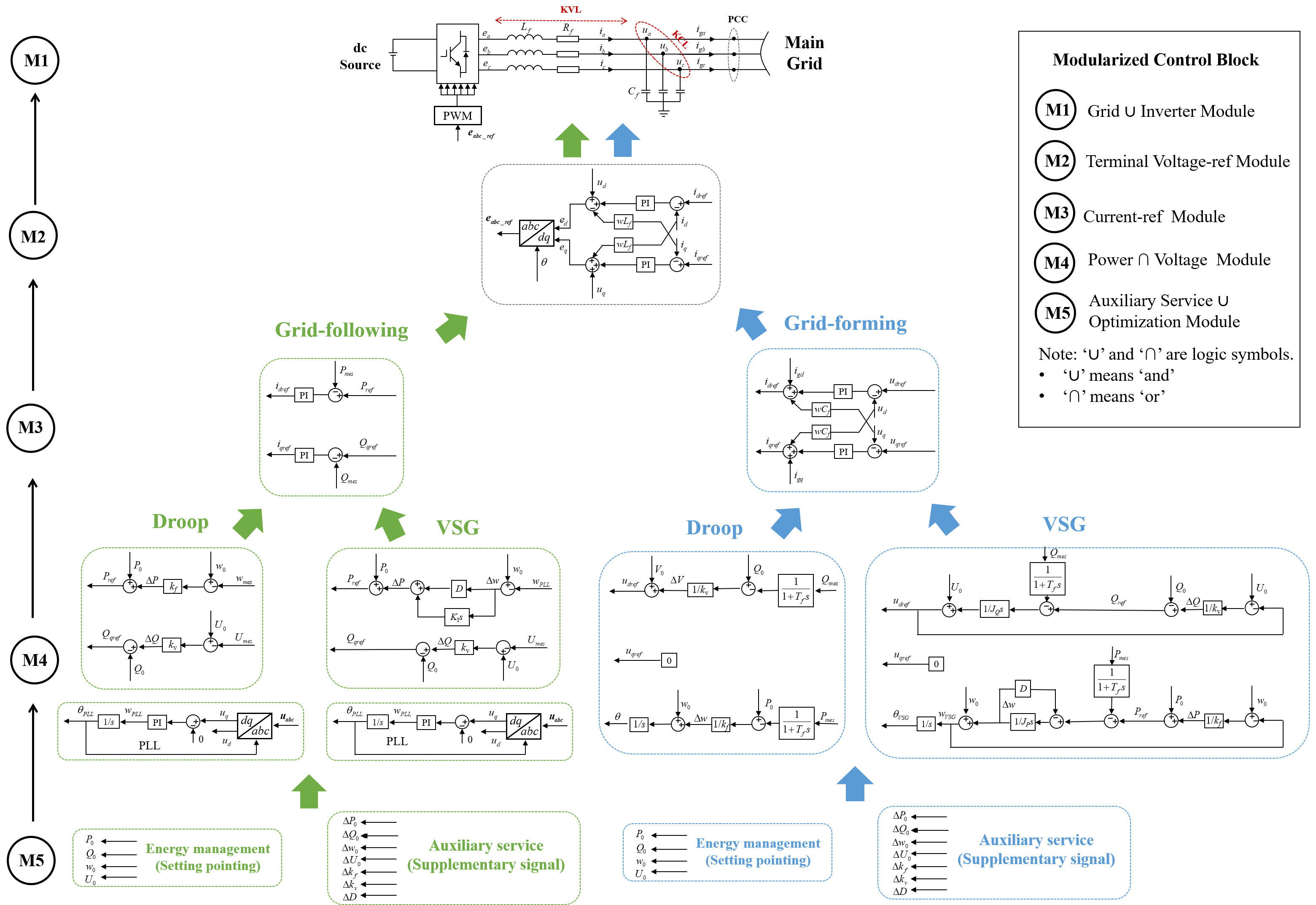}
   \vspace{-0.15cm}
 \caption{Modularized control blocks of GFL and GFM inverters}
 \label{modu_block}
\end{figure*}

 GFL and GFM inverters are no doubt one of the most important units in microgrids \cite{zafar2018design}. This subsection develops the modularized control blocks to present the bottom-level control details of GFL and GFM inverters. Fig. \ref{modu_block} shows the diagram of the modularized control blocks, with which a GFL or GFM inverter can be configured easily by connecting the modules in series. In addition, it is beneficial to the fusion summary in Section IV because the diagram clearly shows the control details that could couple with MFRL.

\subsubsection{M1: Grid $\cup$ inverter module}
The $1^{st}$ module (M1) is named the ‘Grid $\cup$ Inverter Module’ because it illustrates the connection of an inverter to the main grid. As shown in Fig. \ref{modu_block}, the dc source, dc-ac inverter, and RLC filter are linked in series, which are then connected to the main grid through the PCC point. Here, an average model of an inverter that neglects the switching of pulse-width modulation (PWM) is often employed for the control system design. All the high-level controllers work together to generate the reference terminal voltage $e_{abc-ref}$ for PWM.
\subsubsection{M2: Terminal Voltage-ref module}
The $2^{nd}$ module (M2) is named the ‘Terminal Voltage-ref Module’ since it directly generates the reference terminal voltage. The control model is formulated using Kirchhoff's current law (KCL) from $e_{abc}$ to $u_{abc}$ and conducting Park transformation. Then, after implementing proportional-integral (PI) controllers, the physical model and control transfer function in $dq$ framework are shown in (\ref{KCL}) and (\ref{CurRegu_eq}), respectively.
\begin{equation}
L_{f}\left[\begin{array}{c}
\frac{d i_{d}}{d t} \\
\frac{d i_{q}}{d t}
\end{array}\right]+w L_{f}\left[\begin{array}{c}
-i_{q} \\
i_{d}
\end{array}\right]=\left[\begin{array}{c}
e_{d} \\
e_{q}
\end{array}\right]-\left[\begin{array}{l}
u_{d} \\
u_{q}
\end{array}\right]
\label{KCL}
\vspace{-0.25cm}
\end{equation}

\begin{equation}
\begin{aligned}
& {\left[\begin{array}{l}
e_{d} \\
e_{q}
\end{array}\right]=\left[\begin{array}{l}
u_{d} \\
u_{q}
\end{array}\right] } +w L_{f}\left[\begin{array}{c}
-i_{q} \\
i_{d}
\end{array}\right]+ \\
&\left[\begin{array}{cc}
k_{p i d}+\frac{k_{i i d}}{s} & 0 \\
0 & k_{p i q}+\frac{k_{i i q}}{s}
\end{array}\right]
\left(\left[\begin{array}{c}
i_{d r e f} \\
i_{q r e f}
\end{array}\right]-\left[\begin{array}{l}
i_{d} \\
i_{q}
\end{array}\right]\right)
\end{aligned}
\label{CurRegu_eq}
\end{equation}

\subsubsection{M3: Current-ref module}
The $3^{rd}$ module (M3) is named the ‘Current-ref Module’ since it generates the reference current [$i_{dref}$, $i_{qref}$] for M2. For a GFL inverter, [$i_{dref}$, $i_{qref}$] are regulated based on the error between the actual output and the reference value. Eqs. (\ref{M3})-(\ref{Pfiler}) show the transfer function of M3 using PI controllers, where two low-pass filters are used to filter measured power output.
\begin{equation}
\left[\begin{array}{c}
i_{dref} \\
i_{qref}
\end{array}\right]=\left[\begin{array}{cc}
k_{p P}+\frac{k_{i P}}{s} & 0 \\
0 & k_{p Q}+\frac{k_{i Q}}{s}
\end{array}\right]\left[\begin{array}{l}
P-P_{r e f} \\
Q-Q_{r e f}
\end{array}\right]
\label{M3}
\end{equation}

\begin{equation}
\left[\begin{array}{c}
P \\
Q
\end{array}\right]=\left[\begin{array}{cc}
\frac{1}{T_{fP}s+1} & 0 \\
0 & \frac{1}{T_{fQ}s+1}
\end{array}\right]\left[\begin{array}{l}
P_{m e s} \\
Q_{m e s}
\end{array}\right]
\label{Pfiler}
\end{equation}

For a GFM inverter, its physical model is formulated using Kirchhoff's voltage law (KVL) at point $u_{abc}$. After Park transformation and PI controller integration, the algebraic equation and control transfer function in $dq$ framework are shown in (\ref{KVL}) and (\ref{KVL2}), respectively.
\begin{equation}
C_{f}\left[\begin{array}{c}
\frac{d u_{d}}{d t} \\
\frac{d u_{q}}{d t}
\end{array}\right]+w C_{f}\left[\begin{array}{c}
-u_{q} \\
u_{d}
\end{array}\right]=\left[\begin{array}{c}
i_{d} \\
i_{q}
\end{array}\right]-\left[\begin{array}{c}
i_{g d} \\
i_{g q}
\end{array}\right]
\label{KVL}
\end{equation}

\begin{equation}
\begin{aligned}
&{\left[\begin{array}{c}
i_{d r e f} \\
i_{q r e f}
\end{array}\right] } =\left[\begin{array}{c}
i_{g d} \\
i_{g q}
\end{array}\right]+w C_{f}\left[\begin{array}{c}
-u_{q} \\
u_{d}
\end{array}\right]+ \\
&\left[\begin{array}{cc}
k_{p u d}+\frac{k_{i u d}}{s} & 0 \\
0 & k_{p u q}+\frac{k_{i u q}}{s}
\end{array}\right]\left(\left[\begin{array}{l}
u_{dref} \\
u_{qref}
\end{array}\right]-\left[\begin{array}{l}
u_{d} \\
u_{q}
\end{array}\right]\right)
\end{aligned}
\label{KVL2}
\end{equation}

\subsubsection{M4: Power $\cap$ Voltage module}
The $4^{th}$ module (M4) is named the ‘Power $\cap$ Voltage Module’ which indicates the fundamental difference between GFL and GFM inverters. A GFL inverter is controlled as a current source and requires a power reference as an input, while a GFM inverter is controlled as a voltage source and needs a voltage reference as an input \cite{rokrok2018review}. Another big difference is that a GFL inverter needs a PLL to track the phase of the main grid while a GFM inverter is self-synchronized \cite{dinkhah2022optimal}. 
Droop control is the most widely used control method in microgrids. It takes advantage of the coupling between power generation and the grid V-f \cite{qu2020modeling}. Typically, an inductive microgrid employs the $P-f$ and $Q-V$ droop curves, while resistive microgrids uses the reverse $P-V$ and $Q-f$ droop curves. The M4 plotted in Fig. \ref{modu_block} shows the control blocks for an inductive microgrid, and their control models are shown below.

• Droop-controlled GFL inverter
\begin{equation}
\left[\begin{array}{l}
P_{r e f} \\
Q_{r e f}
\end{array}\right]=\left[\begin{array}{ll}
k_{f} & 0 \\
0 & k_{v}
\end{array}\right]\left(\left[\begin{array}{l}
w_{0} \\
U_{0}
\end{array}\right]-\left[\begin{array}{l}
w_{mes} \\
U_{mes}
\end{array}\right]\right)+\left[\begin{array}{l}
P_{0} \\
Q_{0}
\end{array}\right]
\label{DroopGFL}
\end{equation}

• Droop-controlled GFM inverter
\begin{equation}
\left[\begin{array}{c}
w_{ref } \\
u_{dref}
\end{array}\right]=\left[\begin{array}{cc}
\frac{1}{k_{f}} & 0 \\
0 & \frac{1}{k_{v}}
\end{array}\right]\left(\left[\begin{array}{l}
P_{0} \\
Q_{0}
\end{array}\right]-\left[\begin{array}{l}
P_{mes} \\
Q_{mes}
\end{array}\right]\right)+\left[\begin{array}{l}
w_{0} \\
V_{0}
\end{array}\right]
\label{DroopGFM}
\end{equation}

To provide more inertia support to microgrids leveraging DERs, the virtual synchronous generator (VSG) control method is proposed to emulate the behavior of synchronous generators \cite{hou2019improvement}. Mathematically speaking, the VSG belongs to proportional-differential control. Below is the transfer function of the GFL and GFM inverters implementing the VSG.

•	VSG-controlled GFL inverter
\begin{equation}
\begin{aligned}
{\left[\begin{array}{c}
P_{ref} \\
Q_{ref}
\end{array}\right]=} & {\left[\begin{array}{cc}
D+K s & 0 \\
0 & k_{v}
\end{array}\right] \times } \\
& \left ( \left[\begin{array}{l}
w_{0} \\
U_{0}
\end{array}\right]-\left[\begin{array}{l}
w_{mes} \\
U_{mes}
\end{array}\right] \right)
+\left[\begin{array}{l}
P_{0} \\
Q_{0}
\end{array}\right]
\end{aligned}
\end{equation}

•	VSG-controlled GFM inverter
\begin{equation}
\begin{aligned}
{\left[\begin{array}{l}
w_{ref} \\
u_{dref}
\end{array}\right] } = &\left[\begin{array}{cc}
\frac{1}{J_{P} s} & 0 \\
0 & \frac{1}{J_{Q} s}
\end{array}\right] \left \{\left( \left [\begin{array}{l}
P_{ref} \\
Q_{ref}
\end{array}\right]\right.\right.\\
&\left.\left.-\left[\begin{array}{l}
P \\
Q
\end{array}\right]\right)-\left[\begin{array}{l}
D\Delta w \\
0
\end{array}\right] \right\}+\left[\begin{array}{l}
w_{0} \\
U_{0}
\end{array}\right]
\end{aligned}
\end{equation}

Readers are encouraged to check Refs. \cite{zhang2021grid} -\cite{huang2021demand} for some modified VSG and droop control techniques that provide more effective inertia support to microgrids.

\subsubsection{M5: Auxiliary service $\cup$ Optimization module}
Microgrids exploiting M1-M4 can withstand normal disturbances such as load changes and plug-and-play generations. Then, M5 participates in grid optimization and provides auxiliary services, i.e., optimized active and reactive power sharing [28], demand-side management, and V-f support \cite{levron2013optimal}. In order for more economic energy management, M5 also calculates the steady-state setting points such as ($P_0$, $Q_0$) by solving optimal power flow \cite{wang2021adaptive}. On the other hand, it generates the supplementary signals for controller parameters and outputs \cite{de2022balancing} according to the targets of auxiliary service. Review papers regarding M5 can be seen in \cite{bihari2021comprehensive} -\cite{gao2021deep}. 

\subsection{Motivation for MFRL}
\subsubsection{Challenges in the existing control framework}
The high-level research map and modularized control blocks clearly show how existing microgrids are controlled. However, the evolution of microgrids brings more challenges to the existing control framework. The challenges are five-fold: i). The penetration of DERs results in higher uncertainties. Although some robust and stochastic techniques have been employed to address the emerging uncertainties, they are somehow conservative and the probability distribution function still needs to be accurately estimated. ii). It is difficult to model each element of microgrids in detail, i.e., customer behavior. The information that is difficult to model is critical for energy management in M5. iii). Some system parameters are not always accessible; even if accessible, they are not necessarily accurate. iv). Microgrid dynamics are becoming faster because more and more inverter-based resources participate in grid services by adaptively changing their control modes and control parameters. Then, the existing controllers may not be valid anymore. v).Smart grids call for autonomous microgrids, with which engineers and grid operators are free  from parameter tuning for modules in Fig. \ref{modu_block}. Even for other model-free controllers, they still need elaborate tuning for hyper-parameters, i.e., the membership functions of the fuzzy logic controller and the coefficients of the adaption law.

\subsubsection{Why MFRL?}
Microgrid operators have access to massive data sampled by phasor measurement units (PMUs) and advanced metering infrastructures (AMIs) now \cite{chen2021reinforcement}. It opens the possibility for data-driven control. MFRL is an advanced decision-making technique with goal-oriented, data-driven, and model-free characteristics \cite{shuai2021branching}. With the help of MFRL, the uncertainties of the model and parameters may be mitigated through repeated interaction between the environment and the RL agent. It is also beneficial to the autonomous operation of microgrids because the RL agent can actively update its policy based on the microgrid dynamics.

To better fuse MFRL with the existing microgrid control framework, it is necessary to first know the capabilities of each MFRL algorithm, and then choose the proper algorithms in real applications. Thus, the following sections introduce the map of MFRL, the features of main stream MFRL algorithms, and how MFRL can be incorporated into the existing microgrid control framework.

\section{Model-free reinforcement learning}
This section first gives a brief introduction to RL and then summarizes the methodology of MFRL. 
\vspace{-8pt}
\subsection{Formulation of RL}

\begin{figure}[htbp]
   \centering
   \subfloat[\label{RL}]{
        \includegraphics[scale=0.56]{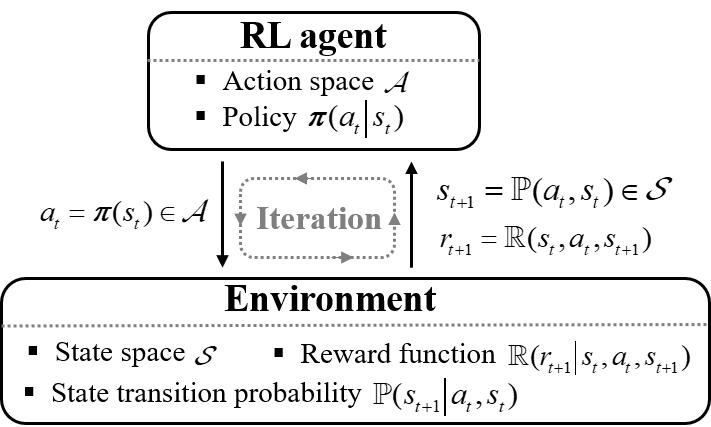}} %
        \vspace{-0.4em} 
   \hfill
   \subfloat[\label{RL_map}]{
        \includegraphics[scale=0.5]{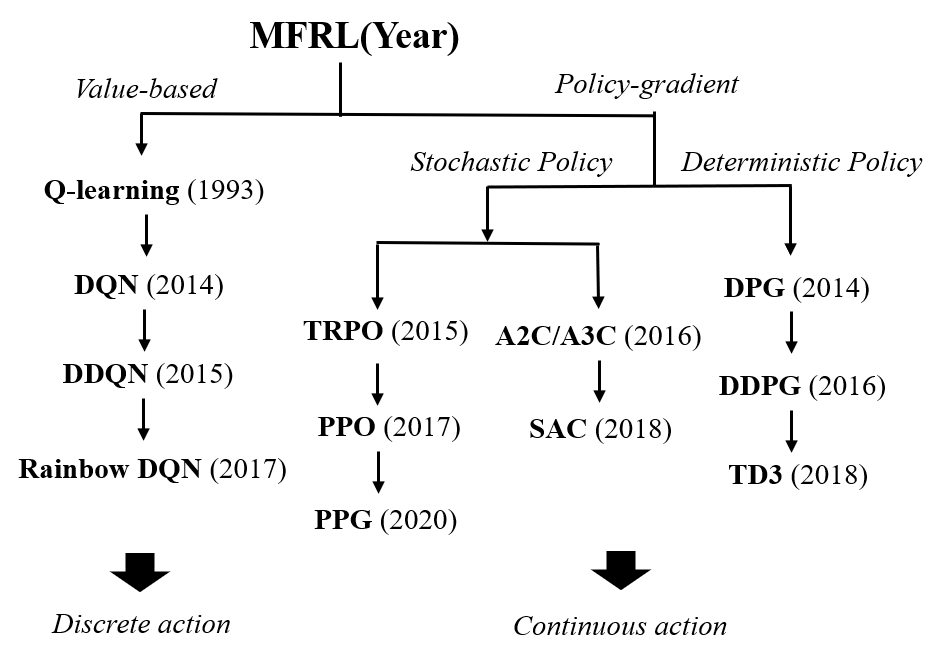}}
   \hfill
\caption{The framework and map of MFRL (a) agent-environment interaction in an MDP (b) methodology}
\vspace{-0.4em} 
\label{RL_diagram}
\end{figure}

RL is a basic ML paradigm formulated as an MDP. As shown in Fig. \ref{RL}, the environment defines the state space $\mathcal{S}$ and the agent holds the action space $\mathcal{A}$. The agent keeps interacting with the environment to update its policy $\pi$ that maps the environment states to actions. In each iteration, the agent chooses action $a_{t} \in A$ according to $\pi$. Then, the environment generates the next state   according to its intrinsic transition probability $\mathbb{P}\left(s_{t+1} \mid s_{t}, a_{t}\right): \mathcal{S} \times \mathcal{A} \rightarrow \Delta(\mathcal{S})$ and feeds back the instant reward $r\left(s_{t}, a_{t}\right)$  to the agent. The iteration is repeated until the agent finds the optimal policy $\pi^{*}$ as follows. 
\begin{equation}
\pi^{*} \in \arg \max _{\pi} J(\pi)=\mathbb{E}_{\pi} \sum_{t=1}^{\infty} \gamma^{t} r\left(s_{t}, a_{t}\right)
\end{equation}
Where $\gamma$ is the discounting factor and $J(\pi)$ is the infinite horizon discounted reward. The optimal policy guarantees the maximum accumulated reward obtained from the environment. 

In MFRL, $\mathcal{A}$ and $\mathcal{S}$ can be either continuous or discrete. For the sake of illustration, this paper uses discrete notation to introduce the methodology.

\subsection{Methodology of MFRL}
Fig. \ref{RL_map} shows the mainstream MFRL methodology. They are categorized into value-based and policy-based algorithms. 

\subsubsection{Value-based algorithms}
The value-based methods learn the \textit{Q}-function that estimates the \textit{Q}-value of state-action pairs $(s, a) \in \mathcal{S} \times \mathcal{A}$. The \textit{Q}-function is denoted as $Q_{\pi}$, based on which the agent can choose the optimal actions with the maximum \textit{Q}-value. According to the Bellman equation,
\begin{equation}
Q_{\pi}\left(s_{t+1}, a_{t+1}\right)=r\left(s_{t}, a_{t}\right)+\gamma E_{s_{t+1}, a_{t+1}} Q_{\pi}\left(s_{t+1}, a_{t+1}\right)
\end{equation}

Through temporal-difference learning, $Q_{\pi}$ can finally converge to its true value under mild assumptions \cite{tsitsiklis1996analysis}.
\begin{equation}
\begin{aligned}
&Q_{\pi}\left(s_{t}, a_{t}\right)=Q_{\pi}\left(s_{t}, a_{t}\right) \\
&\quad+\alpha\left[r_{r}+\gamma \max _{a_{t+1 \in \mathcal{A}}} Q_{\pi}\left(s_{t+1}, a_{t+1}\right)-Q_{\pi}\left(s_{t}, a_{t}\right)\right]
\end{aligned}
\end{equation}

The approximated $Q_{\pi}$ was first recorded in a \textit{Q}-table \cite{tsitsiklis1994asynchronous}. Considering the table’s inefficiency, the deep \textit{Q}-learning network (DQN) \cite{chung2013playing} replaced the \textit{Q}-table with a deep artificial neural network (ANN), which has a strong fitting capability that maps the states to \textit{Q}-value with less memory. Then, the DQN was further improved using the following tricks \cite{van2016deep}:

•	(Prioritized) Reply Buffer enhances the training efficiency. 

•	Double Network relieves the overestimation of \textit{Q}-value. 

•	Dueling Network improves the performance in high-dimensional action space. 

Later, a distributional DQN \cite{bellemare2017distributional} and a quantile regression DQN \cite{dabney2018distributional} were proposed using stochastic policy and distributed training, and they were combined as ‘Rainbow DQN’ by David Silver \cite{hessel2018rainbow} in 2017.

\subsubsection{Policy-based algorithms}

Policy gradient methods directly learn the parameterized policy based on feedback from the environment. Before diving into policy gradient algorithms, it is necessary to introduce the actor-critic (AC) structure. The AC structure has two ANN models that optionally share parameters: i) Critic updates the parameters of value functions; ii) Actor updates the policy parameters under the guidance of the critic. Under the AC structure, policy function can be either stochastic or deterministic. The stochastic policy is modeled as a probability distribution: $a \sim \pi_{\theta}(a \mid s)$, while the deterministic policy is modeled as a deterministic decision: $a=\pi_{\theta}(s)$. They classify the policy-gradient methods.

a) \textit{Stochastic Policy}: As for stochastic policy $a \sim \pi_{\theta}(a \mid s)$, the gradient of the expected reward to policy parameters is calculated according to policy gradient theorem \cite{sutton1999policy} as follows
\begin{equation}
\nabla J(\theta)=\sum_{s \in \mathcal{S}} \mu_{\theta}(s)
\sum_{a \in \mathcal{A}} \pi_{\theta}(a \mid s) Q_{\pi_{\theta}}(s, a) \nabla_{\theta} \ln \pi_{\theta}(a \mid s)
\end{equation}
Where $\mu_{\theta}(\mathcal{S}) \in \Delta(\mathcal{S})$ is the state distribution. Then, the policy is updated using the gradient ascent method
\begin{equation}
\theta_{t+1}=\theta_{t}+\eta \nabla J\left(\theta_{t}\right)
\end{equation}
Where $\eta$ is the learning rate. It is necessary to avoid large updating of step size in each iteration since the policy gradient readily falls into a local maximum. To make the policy gradient training more stable, trust region policy optimization (TRPO) added a Kullback–Leibler (KL) divergence constraint to the process of policy updating \cite{schulman2015trust}. It solves the optimization problem as follows
\begin{equation}
\begin{aligned}
\max _{\theta} & J(\theta)=\mathbb{E}\left[\frac{\pi_{\theta}^{\prime}(a \mid s)}{\pi_{\theta}(a \mid s)} \hat{A}_{\theta}(a \mid s)\right] \\
& s.t.  \mathbb{E}\left[D_{K L}\left(\pi_{\theta}{}^{\prime} \|\pi_{\theta}\right)\right] \leq \delta
\end{aligned}
\end{equation}
Where $\pi_{\theta}^{\prime}$ is the new policy; $D_{KL}$ is the KL-divergence. 

Considering the complexity of measuring  $D_{KL}$ in each update, proximal policy optimization (PPO) was developed to accelerate the training \cite{schulman2017proximal}. PPO uses a clipped surrogate objective while retaining similar performance as follows
\begin{equation}
\begin{aligned}
\max _{\theta} & J(\theta)=\mathbb{E}\left\{\operatorname { m i n } \left[\frac{\pi_{\theta}{ }^{\prime}(a \mid s)}{\pi_{\theta}(a \mid s)} \hat{A}_{\theta}(a \mid s)\right.\right.\\
&\left.\left.\operatorname{clip}\left(\frac{\pi_{\theta}{ }^{\prime}(a \mid s)}{\pi_{\theta}(a \mid s)}, 1-\varepsilon, 1+\varepsilon\right) \hat{A}_{\theta}(a \mid s)\right]\right\}
\end{aligned}
\end{equation}

In PPO, the actor network and critic network share the same learned features, and this may result in conflicts between competing objectives and simultaneous training. Hence, a phasic policy gradient (PPG) separates the training phased for actor and critic networks \cite{cobbe2021phasic}, which leads to a significant improvement in sampling efficiency.
Other improved versions of the AC structure include advantage actor-critic (A2C), asynchronous advantage actor-critic (A3C), and soft actor-critic (SAC). A2C and A3C both enable parallel training using multiple actors, but the actors of A2C work synchronously, and those of A3C work asynchronously \cite{mnih2016asynchronous}. SAC improves the exploration of agents incorporating policy entropy \cite{haarnoja2018soft}.

b) \textit{Deterministic Policy}: The gradient of deterministic policy $a=\pi_{\theta}(s)$ is expressed as
\begin{equation}
\nabla J(\theta)=\left.\mathbb{E}_{s \sim \mu_{\theta}} \nabla_{a} Q_{\pi_{\theta}}(s, a)\right|_{a=\pi_{\theta}(s)} \nabla_{\theta} \pi_{\theta}(s)
\end{equation}

The deterministic policy gradient (DPG) method firstly used deterministic policy \cite{silver2014deterministic}. Then, the deep deterministic policy gradient (DDPG) was developed by combining the DPG and DQN \cite{lillicrap2015continuous}. The DDPG extends the discrete action space of the DQN to continuous space while learning a deterministic policy. Later, the twin delayed deep deterministic (TD3) policy gradient applied three tricks, i.e., clipped network, delayed update of critic network, and target policy smoothing to prevent the overestimation of \textit{Q}-value in the DDPG. 

\subsubsection{Summary}
The DQN, DDPG, and A3C are three basic paradigms of MFRL representing value-based methods, deterministic policy methods, and stochastic policy methods. Their upgraded versions, the Rainbow DQN, TD3, and PPG, SAC represent the state-of-the-art of each paradigm, which are the best choices for fusing MFRL with the existing microgrid control framework. Moreover, the value-based methods such as DQN are more suitable for discrete control tasks like transformer tap and switch on/off control, while the policy-based methods like TD3 are more suitable for continuous tasks such as active power and reactive power reference generation.

\section{Fusion of model-free reinforcement learning with microgrid control}
Section II and Section III introduce the existing microgrid control framework and the MFRL, separately. This section furthers the fusion details, including the application guidelines and the challenges and insights of using MFRL in microgrid control.

\subsection{Application guideline}

\subsubsection{Problem formulation}

Microgrid control is intrinsic to an infinite MDP that MFRL can solve. Ref. \cite{chen2022reinforcement} answered the question of ‘\textit{How}’, that is, ‘\textit{How to formulate a control problem that can be solved by MFRL?}’, which includes four steps: i). Determine the environment, state space $\mathcal{S}$, and action space $\mathcal{A}$; ii) Design reward function $\mathcal{R}$ according to control targets; iii). Select proper learning algorithm; iv). Train agent and validate the learned policy. The four steps are exemplified below based on two specific application scenarios, frequency regulation and voltage regulation.

i) Formulation of frequency regulation: Eqs. (\ref{Sf})-(\ref{Rf}) show the general configuration of a MFRL agent for frequency regulation in microgrids. The agent has unique action space when fusing with different modules in Fig. \ref{modu_block}.
\begin{equation}
\mathcal{S}_f:=\left[\left(w_i\right)_{i \in \mathcal{N}},\left(P_{i j}, Q_{i j}\right)_{i j \in \varepsilon}\right]
\label{Sf}
\end{equation}

\begin{equation}
\mathcal{A}_f=\left\{\begin{array}{l}
\mathrm{M2}:\left[\left(e_{a b c, i}\right)_{i \in \mathcal{I}}\right] \\
\mathrm{M3}:\left[\left(i_{d, i}, i_{q, i}\right)_{i \in \mathcal{I}}\right] \\
\mathrm{M4}:\left[\left(P_{r e f, i}, Q_{r e f, i}\right)_{i \in \mathcal{I}_{\text {GFL }}},\right. \\
\left.\quad\left(w_{r e f, i}, u_{d r e f, i}\right)_{i \in \mathcal{I}_{G F M}}\right] \\
\mathrm{M5}:\left[\left(P_{0, i}, Q_{0, i}, \Delta x_i\right)_{i \in \mathcal{I}}\right]
\end{array}\right.
\end{equation}

\begin{equation}
\mathcal{R}_f(t)=-\sum_{i \in \mathcal{N}}\left[w_i(t)-w_0\right]^2
\label{Rf}
\end{equation}
Where $w_i$ is frequency at each bus $i$; ($P_{i j}, Q_{i j}$) is the power flow over line from bus $i$ to bus $j$; M2-M5 are the modules summarized in Fig. 2; $\mathcal{I}$ is the inverter set; $\mathcal{I}_{GFL}$ and $\mathcal{I}_{GFM}$ are the set of GFL inverters and GFM inverters, respectively. Since the control target is to maintain frequency, the deviation of frequency is designed as reward function.

ii) Formulation of voltage regulation: Eqs. (\ref{Sv})-(\ref{Rv}) show the general configuration of a MFRL agent for frequency regulation in microgrids.

\begin{equation}
\mathcal{S}_v:=\left[\left(v_i\right)_{i \in \mathcal{N}},\left(P_{i j}, Q_{i j}\right)_{i j \in \varepsilon},\left(\tau_i\right)_{i \in \mathcal{T}}\right]
\label{Sv}
\end{equation}
\begin{equation}
\mathcal{A}_v=\left\{\begin{array}{l}
\mathrm{M}_2:\left[\left(e_{a b c, i}\right)_{i \in \mathcal{I}}\right] \\
\mathrm{M}_3:\left[\left(i_{d, i}, i_{q, i}\right)_{i \in \mathcal{I}}\right] \\
\mathrm{M}_4:\left[\left(P_{r e f, i}, Q_{r e f, i}\right)_{i \in \mathcal{I}_{G F L}},\right. \\
\left.\quad\left(w_{r e f, i}, u_{d r e f, i}\right)_{i \in \mathcal{I}_{G F M}}\right] \\
\mathrm{M}_5:\left[\left(P_{0, i}, Q_{0, i}, \Delta x_i\right)_{i \in \mathcal{I}},\left(\tau_i\right)_{i \in \mathcal{T}}\right]
\end{array}\right.
\end{equation}

\begin{equation}
\mathcal{R}_v(t)=-\sum_{i \in \mathcal{N}}\left[v_i(t)-v_0\right]^2
\label{Rv}
\end{equation}
Where $v_i$ is the voltage magnitude of bus $i$, and $\tau_i$ is the tap positions of the on-load tap changers (OLTPs) of transformers. Compared with frequency regulation, the agent has distinct action of OLTPs in M5 for voltage regulation.

After selecting $\mathcal{S}$, $\mathcal{A}$, and $\mathcal{R}$, the mainstream MFRL algorithms are selected to update the policy of the agent. Note that the selected algorithms should be applicable to the application scenarios. For example, the discrete algorithm in Fig. \ref{RL_map} is better for discrete control actions like OLTPs. In addition, the above formulations give a general form of configurating an MFRL agent for microgrid control, and they can be modified according to customized control tasks.

In addition to problem formulation, there are another two fundamental questions regarding ‘What’ that remain to be answered. They are

• \textit{Q1: What kinds of tasks is MFRL suitable for? }

• \textit{Q2: How can MFRL be fused with the existing microgrid control framework? }

The following two subsections tries to answer these two questions based on the state-of-the-art of MFRL. The answers can serve as the application guideline for adopting MFRL in microgrids.

\subsubsection{\textit{What kinds of tasks is MFRL suitable for?}}
In general, MFRL is suitable for tasks with the following four features: 

i) Relatively unchanged environment. Policy learned by RL agents reflects the physical law in the training environments, which fundamentally determines the state transition probability. As shown in the diagram in Fig. \ref{RL}, environment generates rewards based on $\mathbb{P}\left(s_{t+1} \mid s_{t}, a_{t}\right): \mathcal{S} \times \mathcal{A} \rightarrow \Delta(\mathcal{S})$ and feed the rewards to RL agent for policy updating. A new environment has distinct state transition probability function, which may have conflicts with the buffer data and trained policy. Thus, the working environment should not differ too much from the training environment. That's why in Tab. 1, the training microgrids and validation microgrids usually have fixed topology and predefined disturbances.

ii) Clear control target. Clear control targets facilitate the design of reward functions. The objective function in the optimization problem, optimal control, and MPC can be directly transformed to a reward function. With the function grouping and hierarchical structure in Fig. 1, the specific control targets can be briefly categorized into frequency regulation, voltage regulation equation, and economic benefits. Then, the voltage deviation \cite{chen2021powernet}, frequency deviation \cite{gheisarnejad2020novel} and energy management cost/revenue \cite{samende2022multi} -\cite{shuai2020online} are transformed into reward functions in (\ref{Rf}) and (\ref{Rv}). Crucially, a well-designed reward function gives the MFRL agent the best guide to learn the optimal policy. 

iii) Available data. Environmental data must be accessible if the agent interacts with a real system. Also, the real environment should tolerate improper actions for exploration. If the environment is a simulator, the simulation should run quickly to allow for thousands of repetitions. For example, a fast a simulator and a real tokamak vessel were developed for training and validation in \cite{degrave2022magnetic}.

iv) Acceptable control complexity. ‘Acceptable’ means the control complexity should be neither too low nor too high. For each perspective summarized in the high-level research map, there is no research trying to replace all the controllers. Most of the research just focused on a specific task that a model-based controller cannot handle but MFRL can, because there is no need to replace a simple model-based controller that has good performance and it is unrealistic to let AI directly control the whole microgrid for now.

\subsubsection{\textit{How can MFRL be fused with the existing microgrid control framework?}}
MFRL is essentially a useful tool that serves microgrid control. It follows microgrid control targets when fused with the existing control framework. In general, there are three ways of fusing as follows.

i). Model identification and parameter tuning. MFRL assists in identifying the uncertain models of the grid components accurately. Also, it can address the uncertainty and unavailability of model parameters and release the grid operators from complex and time-consuming parameter tuning, especially tuning a large model with many parameters.

ii). Supplementary signal generation. MFRL can generate the supplementary control signals for model-based controllers, with which the current controllers can be made more robust and deal with complicated control tasks.

iii). Controller substitution. MFRL can completely replace the existing model-based controllers if they are no longer effective. It needs fewer inputs but has better performance than model-based controllers owing to the ANN’s strong fitting capability.

In general, the application guide is summarized based on the existing microgrid control research that employ MFRL. The detailed literature review will be performed in the next subsection.

\subsection{Literature review}
Sorted in the way of fusing, Table \ref{RL_literature_review} summarizes the literature adopting MFRL in microgrids, where the key features are listed in the last column. In general, MFRL has fused with the optimization and control tasks in microgrids. Most research has tried to replace the existing model-based controllers with MFRL agents. In addition, more researchers focus on optimization problems that have clear targets. The objective functions are directly transformed or incorporated into the reward function.

\begin{table*}
 \caption{Literature summary of implementing MFRL in microgrids}
\label{RL_literature_review}
\begin{tabularx}{\textwidth}{llp{1.8cm}p{1.5cm}p{2.1cm}p{2.5cm}p{6.5cm}}
\toprule
\textbf{Ref.}     &\textbf{Year}    & \textbf{Topic}      & \textbf{Algorithm}   &\textbf{Way of fusing}      &\textbf{Environment}      & \textbf{Key features} \\
\hline
\addlinespace

\cite{arwa2020reinforcement}      & 2020     &Converter voltage stability & PPO & Parameter tuning for M4 & dSPACE MicroLabBox  & 1) Adaptively tune the feedback gains of the ultra-local model; 2) Mitigate the stability issues caused by constant power loads \\
\addlinespace

\cite{abianeh2021vulnerability}     &2022 &Cyber attack & DQN, Multi-agent DDPG &Model identification for M2-M4 &MATLAB/Simulink, dSPACE MicroLabBox DS1202 & 1) RL agent automatically discover the vulnerable spots and generate coordinated destabilizing false data attacks; 2) Enhanced agent with a sniffing feature to enable maintaining the stealthy attacks under connection failure.\\
\addlinespace

\cite{wang2020multi}      & 2021     &Transient stability  &SAC  & Model identification  & Numerical simulator & 1) Consider multiple events that result in transient stability simultaneously; 2) Test learned policy in new events \\
\addlinespace

\cite{neal2021reinforcement}     &2020     & Microgrid Penetration Test   &A3C  & Supplementary signal generation for M5 & Numerical simulator & 1) Perform Penetration Testing for microgrids 2) RL agent uncovers the malicious input that can compromise the effectiveness of the controller \\
\addlinespace

\cite{hajihosseini2020dc}     & 2021 & dc-dc buck converter control &DDPG &Supplementary signal generation for M4 & dSPACE MicroLabBox and DS1302 I/O board & 1) Design an intelligent PI controller based on sliding mode observer to mitigate instability; 2) RL agent generates the auxiliary signals to reduce the error of observer\\
\addlinespace

\cite{gheisarnejad2020novel}     &2021 &Secondary frequency control &DDPG &Supplementary signal generation for M5 &Matlab and dSPACE 1202 board &1) Consider Type-II fuzzy system; 2) Generate supplementary signals for PI-based secondary controllers\\
\addlinespace

\cite{adibi2022secondary}     &2022 &Secondary frequency control & A2C &Supplementary signal generation for M4 &Discrete-time ODE model in numerical simulator &1) Achieve frequency synchronization within an ultimate bound given dominantly resistive and/or inductive line and load impedances; 2) A feedback control is formulated based on the unknown dynamics, using Lyapunov theory.\\
\addlinespace

\cite{li2021reinforcement}     &2021 &Mode transition & Q-learning &Supplementary signal generation for M4 &Numerical simulator &1) RL agent generates voltage angle and magnitude adjustments; 2) Enables bulk power grid restoration by using microgrids' black start capability.\\
\addlinespace

\cite{du2019intelligent}     &2019 &Energy management &Q-learning & Controller substitution in M5 & Matlab and Python & 1) Perform privacy-preserved response learning for multi-microgrids; 2) Implement Monte Carol method for decision making\\
\addlinespace

\cite{yan2019data}     &2019 &Battery SOC control &DDPG &Controller substitution in M4 &Numerical simulator &1) Perform supervised pre-training for critic-network based on control cost; 2) Perform pre-training for actor-network based on the output of PI controllers
\\
\addlinespace

\cite{duan2019reinforcement}     &2019 &Energy storage system control &Q-learning &Controller substitution in M4 and M5 &Matlab-Simulink Simscape toolbox &1) Optimize the charging and discharging profile to suppress the disturbance caused by integrating a new hybrid energy system; 2) One network estimates the unknown system dynamics and the other solves the optimal policy \\
\addlinespace

\cite{elsayed2019low}     &2020 &Emergency control &DQN &Controller substitution in M5 &InterPSS in java and OpenAI in python & 1) Train RL agent under the circumstance of predefined topology and random short-circuit faults; 2) Use hybrid simulation with java and python\\
\addlinespace

\cite{wang2021unintentional}     &2021 &Islanding transition control &Q-learning &Controller substitution in M3-M5 &Matlab-Simulink &1) Update specific values or parameters in reinforcement learning with artificial emotion; 2) Implement load shedding to reduce the impacts of intentional islanding\\
\addlinespace

\cite{wang2021unintentional}     &2021 &Energy storage system control &DDQN &  Controller substitution in M4  & Numerical simulator (TensorFlow and GUROBI) &1) Improve robustness with prioritized replay policy based on sum-tree; 2) RL agent directly outputs actions without solving a model-based optimization problem \\
\addlinespace

\cite{khooban2020novel}     &2022 &Peer-to-peer energy trading &Multi-agent TD3 & Controller substitution in M5 &Numerical simulator & 1) Consider both external peer-to-peer energy trading and internal energy conversion; 2) The high-dimensional decision-making problem is solved by multi-agent TD3 under the resolution of hours\\
\addlinespace

\cite{chen2021powernet}     & 2021 & Secondary voltage control &  Multi-agent A2C  & Controller substitution in M5  & Numerical simulator  & 1) A high-fidelity powergrid simulation platform 'PGSim' is developed as environment; 2) Enhance the scalability with spatial discount factor to reduce the effect from remote agents.\\
\addlinespace

\cite{samende2022multi}     & 2022  & P2P trading & Multi-agent DDPG & Controller substitution in M5 &  Numerical simulator & 1) Incentivize energy trading with distribution network tariffs to satisfy the physical constraints; 2) Do simulations based on real-world trading data sets. \\
\addlinespace

\cite{xiong2022two}     & 2022 & Energy management & Trust region PPO & Controller substitution in M5 & Numerical simulator & 1) The distribution system operator (DSO) is viewed as a RL agent without knowing user information for privacy protection; 2) Integrate a differentiable trust region layer to improve the robustness of the policy updating.\\
\addlinespace

\cite{hao2021distributed}     & 2021 & Online economic dispatch & Hierarchical Q-learning & Controller substitution in M5  & Numerical simulator & 1) Offers a subtle blend of immediate and future rewards to guarantee a long-term performance; 2) Integrate domain knowledge to narrow down learning space to a feasible region and avoids violations.\\
\addlinespace

\bottomrule
\end{tabularx}
\vspace{-0.25cm}
\end{table*}

\subsection{Challenges and insights}
Although many researchers have been investigating the applications of MFRL in microgrid control, there is still a clear gap between theory (simulation) and practice (real microgrid operation). The main concerns are the aspects of the environment, scalability, generalization, security, and stability. This subsection summarizes these challenges and gives some insights on how to tackle them.

\subsubsection{Environment}

\begin{figure}[htbp]
 \centering
   \vspace{-0.25cm}
   \includegraphics[scale=0.7]{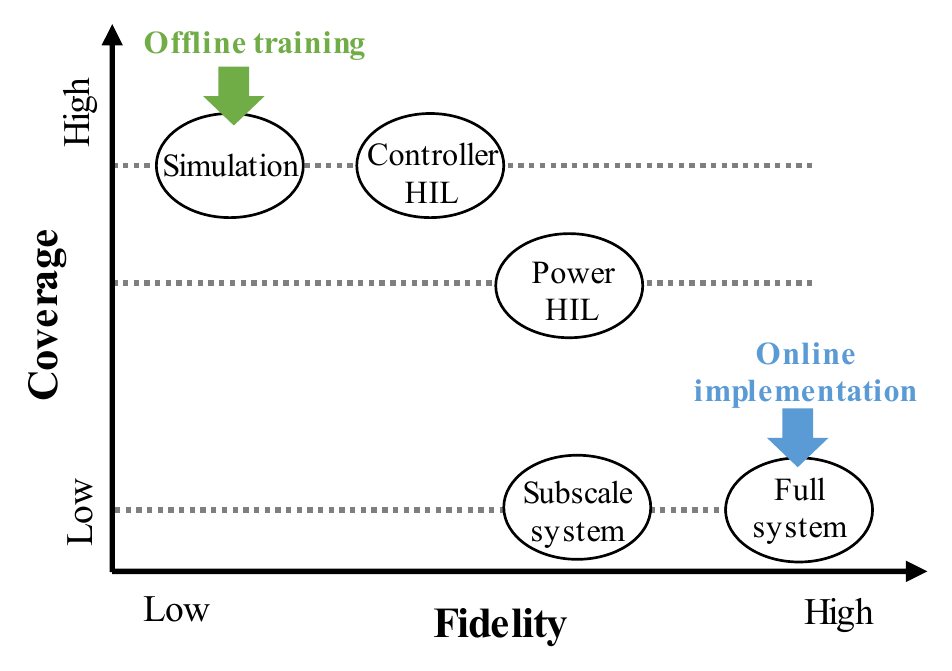}
   \vspace{-0.25cm}
 \caption{Microgrid testbeds \cite{ieee2017ieee} and MFRL environment}
 \vspace{-0.35cm}
 \label{testbed}
\end{figure}

•	Challenges: As shown in Fig. \ref{testbed}, the conventional model-based microgrid controllers have several types of tests before implementation, i.e., simulation, controller hardware in the loop (HIL) test, power HIL test, subscale system test, and full system test. They are the options for the MFRL environment. Existing literature suggests offline training in the numerical simulator and online implementation in real systems \cite{duan2019reinforcement} because the RL agent requires sufficient exploration during training which is unrealistic in HIL or real systems. That’s why early RL was mainly used in video games, where the simulator could perfectly emulate the working environments \cite{hao2021distributed}. Among the current power testbed types, simulation has the highest coverage of test scenarios but the least fidelity, which is the major concern of employing MFRL. Even if the agent learned a good policy in a numerical simulator, it may not function effectively in a real microgrid.

•	Insights: As for numerical simulators, they are on the way to developing a more accurate and faster toolbox capable of serving as a high-fidelity MFRL environment. Improved power system modeling \cite{hoidalen2021analysis} and more efficient numerical simulation techniques, such as the hybrid symbolic-numeric framework \cite{shi2018analytical}, are currently being developed. Further, it would be better to develop a standardized and customized training environment that assists in setting up the interface with power simulators such as PSCAD, PSSE, and MATLAB-Simulink, just like “Gym” in the field of deep RL \cite{cui2020hybrid}. The standardized environment can also serve as a baseline for algorithm tests and comparisons. On the other hand, it is a good way to design a HIL test system that is equipped with specialized protection and can tolerate random exploration to some degree. In this way, the HIL test system may work as an environment that closely resembles an actual microgrid. 
Moreover, MFRL agent can learn from historical data. To improve the learning efficiency and address the problem of real-data insufficiency, some advanced techniques have been developed. For example, i) long-tail learning \cite{liu2020deep} can learn effectively on biased data set; ii). deep active learning \cite{zhang2022deep} can also be used to more efficiently label disturbance data.

\subsubsection{Scalability}
•	Challenges: MFRL suffers from the curse of dimensionality like some model-based controllers. The expansion of state space and action space will result in an exponential increase in control complexity, thereby increasing the difficulty of exploration and training. Existing MFRL research on microgrid control mainly focuses on some small-scale problems \cite{song2021prioritized} and utilizes ANN with a few layers. To promote the application of MFRL in microgrid control, it is necessary to improve its scalability.

•	Insights: On the one hand, it is an effective way to reduce control complexity by integrating domain knowledge into problem formulation. For example, \cite{charles2019curse} narrowed down the learning space and avoided baseline violations based on the generation constraints. On the other hand, it would be better to increase the capability of existing MFRL models by: i). increasing the exploration efficiency by designing guided exploration strategies like evolutionary RL \cite{khadka2018evolutionary}; ii). increasing the fitting capability of ANN through the modern design of network structures, i.e., sequential-to-sequential networks and transformers \cite{vaswani2017attention}; iii). increasing the training efficiency through distributed techniques like federated learning \cite{li2021survey} and edge computing \cite{cao2020overview}. All of these methods can help relieve the pressure on training and make MFRL more scalable for microgrid control.

\subsubsection{Generalization}
•	Challenges:	Similar to DL, MFRL was accused of “inability of generalization” because a well-trained agent does not function effectively in a changing environment \cite{agarwal2021contrastive}. Even in an unchanged environment, the diversity of disturbances may also distort the agent. In microgrid control, it is difficult to cover all the disturbances during the training, which is critical on the condition that RL agents replace the existing controllers. 

•	Insights: Firstly, rich training scenarios benefit the generalization of MFRL. For example, \cite{liu2020two} addressed the uncertainty of Volt-Var control in active distribution systems by generating a bunch of offline training scenarios. It is also a good way to employ robust RL that can tolerate the uncertainty of the environment \cite{pinto2017robust}. Further, transfer learning can also enhance the MFRL’s generalization capability, which has proven to be effective in the field of DL.

\subsubsection{Security}
•	Challenges: Security is referred to as static security in this paper, meaning that system state should respect the static physical constraints to avoid damaging the device. In microgrids, these constraints can be thermal limit constraints and control signal constraints decided by the physical capability of microgrid components. They are usually explicit and known according to microgrid device manufacture, and there are IEEE Standards setting the secure operational range of voltage and frequency. However, due to the non-interpretability of ANN, the learned policy cannot always guarantee each variable respect the constraints. Furthermore, it is also a problem to guarantee secure exploration in a HIL or real system. In the future, MFRL agents may be trained in a HIL microgrid to overcome the shortcomings of numerical simulators, where the exploration cannot violate the physical constraints of the HIL or real system for sure.

•	Insights: Through constrained RL \cite{brunke2022safe} - \cite{junges2016safety} and safe RL \cite{li2021safe} - \cite{berkenkamp2017safe}, the actions of RL agents can be projected to a safety region and thus always respect the physical operational constraints. In addition, physics-constrained and physics-informed deep learning \cite{huang2022applications} is also under development and can be integrated into MFRL to address security concerns. In physics-constrained deep learning, a “safety layer” is often leveraged to maintain constraint satisfaction under different physics knowledge, while physics-informed learning embeds the knowledge of physical laws that govern by partial differential equations into training.

\subsubsection{Stability}
•	Challenges: Stability is referred to as dynamic stability under a disturbance. According to the definition in \cite{hatziargyriou2020definition}, the stability is the ability of an electric power system, for a given initial operating condition, to regain a state of operating equilibrium after being subjected to a physical disturbance, with most system variables bounded so that practically the entire system remains intact. Model-based microgrid controllers must pass the stability test through eigenvalue analysis or the Lyapunov function validation before implementation. However, the employment of MFRL challenges the model-based criteria because the uninterpretable RL agents dramatically change the closed-loop dynamics of microgrids.

•	Insights: Integrating domain knowledge is the best way to guarantee microgrid stability for now. As for the first two fusing approaches, i) model identification and parameter tuning and ii) supplementary signal generation, 
model-based stability criteria can still be used to verify the system stability because the MFRL agent doesn’t break down closed-loop systems. MFRL complements the model-based approaches and improves them in a data-driven way. The supplementary signals generated by the MFRL agent can be viewed as hyper-parameters. Through techniques like semi-definite programming (SDP), linear matrix inequality (LMI), and sum-of-square programming \cite{liu2022ensuring}, the security range of these hyper-parameters can be obtained to guarantee dynamic stability \cite{zhang2021domain}. As for the third way of fusion, the complete controller substitution, MFRL agents dramatically change the closed-loop dynamics and make the system difficult to model. To address the stability issues in this condition, this paper gives three potential solutions. i). enrich the training data and training scenarios. The learned policies basically reflect the state transition of the environment. If the training data set has covered sufficient instability scenarios, the corresponding punishment reward can help RL agents avoid unstable actions. ii). use a physics-informed approach by integrating model-based stability criteria into MFRL training. For example, the Lyapunov function \cite{chow2018lyapunov} and the Gaussian process estimation  \cite{berkenkamp2017safe} can be used to generate stability criteria for MFRL training, and \cite{cui2021lyapunov} proposed a Lyapunov-regularized RL for transmission system transient stability. iii).perform policy stability validation through time-domain simulation (TDS). TDS has been widely used in power systems to validate the stability of nonlinear components or modules. It can also help validate the stability of the inexplicable RL policy.

\section{Conclusion}
Model-based controllers are still the foundation of existing microgrid control systems. However, the emerging challenges caused by the uncertainty of DERs and extreme weather call for advanced control techniques. As a model-free and data-driven approach, MFRL opens the possibility of non-linear, high-dimensional, and high-complex microgrid control. It may contribute to a huge upgrade of the existing control framework. 

Against this background, this paper firstly performs a comprehensive review of the current microgrid control framework and then summarizes the applications of MFRL. In general, there are three ways of fusing MFRL with the existing model-based controllers, including i). model identification and parameter tuning, ii). supplementary signal generation, and iii). controller substitution. For now, there is still an obvious gap between the theory (simulation) and its practical application. The challenges are mainly categorized into environment, scalability, generalization, security, and stability. With the rapidly developed techniques in the fields of both power and artificial intelligence, the author believes the challenges summarized in this paper will finally be overcome. Someday in the future, the MFRL can perfectly fuse with the existing microgrid control framework.

\section*{Acknowledgment}
The authors would like to thank the financial support in part from the US DOD ESTCP program under the grant number EW20-5331 to complete this research work.

\bibliographystyle{IEEEtran}
\bibliography{bib_bshe} 

\begin{IEEEbiography}
[{\includegraphics[width=1in,height=1.6in, clip, keepaspectratio]{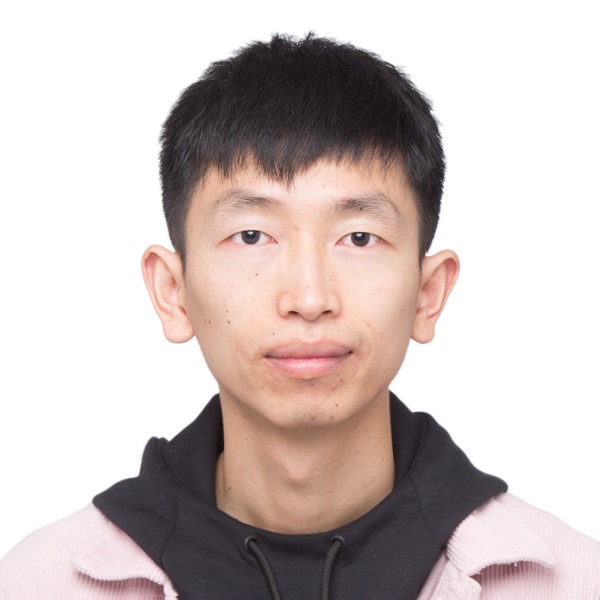}}]{Buxin She}
(Student Member, IEEE) is presently a Ph.D. student in the department of electrical engineering and computer science at The University of Tennessee, Knoxville. He received his B.S.E.E. and M.S.E.E. degrees both from Tianjin University, Tianjin, China in 2017 and 2019, respectively. His research interests include microgrid operation and control, machine learning in power system, and power grid resilience.
\end{IEEEbiography}

\begin{IEEEbiography}
[{\includegraphics[width=1.2in,height=1.2in, clip, keepaspectratio]{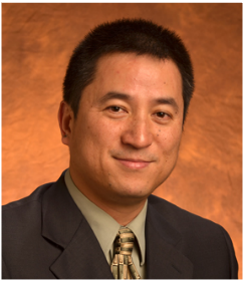}}]{Fangxing Li}
(Fellow, IEEE) is also known as Fran Li. He received the B.S.E.E. and M.S.E.E. degrees from Southeast University, Nanjing, China, in 1994 and 1997, respectively, and the Ph.D. degree from Virginia Tech, Blacksburg, VA, USA, in 2001. Currently, he is the James W. McConnell Professor in electrical engineering and the Campus Director of CURENT at the University of Tennessee, Knoxville, TN, USA. His current research interests include resilience, artificial intelligence in power, demand response, distributed generation and microgrid, and energy markets. From 2020 to 2021, he served as the Chair of IEEE PES Power System Operation, Planning and Economics (PSOPE) Committee. He has been serving as the Chair of IEEE WG on Machine Learning for Power Systems since 2019 and the Editor-In-Chief of IEEE Open Access Journal of Power and Energy (OAJPE) since 2020.

Dr. Li has received numerous awards and honors including R\&D 100 Award in 2020, IEEE PES Technical Committee Prize Paper award in 2019, 5 best or prize paper awards at international journals, and 6 best papers/posters at international conferences.
\end{IEEEbiography}

\begin{IEEEbiography}
[{\includegraphics[width=1.05in,height=2in, clip, keepaspectratio]{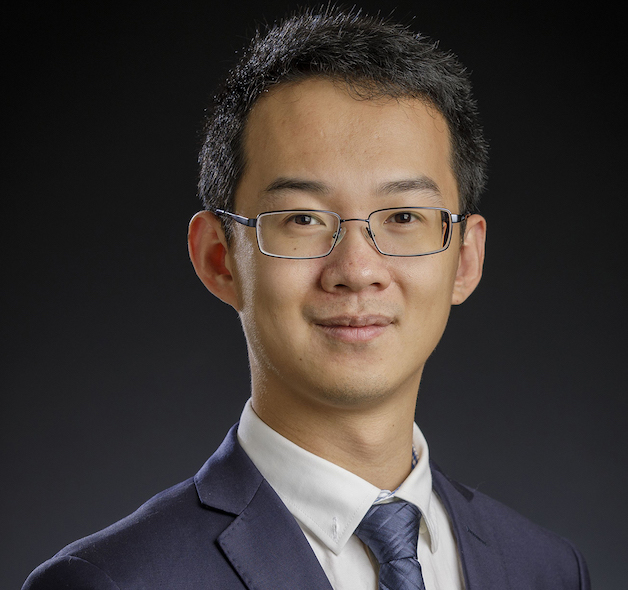}}]{Hantao Cui}
Hantao Cui (Senior Member, IEEE) received the B.S. and M.S. degrees from Southeast University, China in 2011 and 2013, and the Ph.D. degree from the University of Tennessee, Knoxville in 2018, all in electrical engineering. He is currently an Assistant Professor with the School of Electrical and Computer Engineering, Oklahoma State University. His research interests include power system modeling, simulation, and high-performance computing.
\end{IEEEbiography}

\begin{IEEEbiography}
[{\includegraphics[width=1.1in,height=1.1in, clip, keepaspectratio]{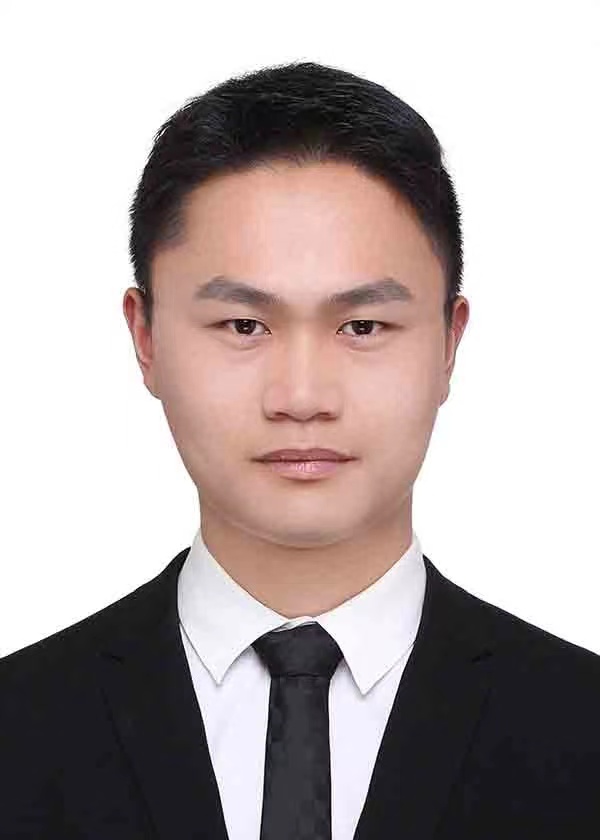}}]{Jinqiu Zhang}
(Student Member, IEEE) received the B.S. and M.S. degree in electrical engineering from Tianjin University, Tianjin, China in 2016 and 2019, respectively. He is currently
working towards his Ph.D. degree in electrical and
computer engineering at the National University of
Singapore, Singapore. His current research interests include cyber security of power grids, distributed control and optimization in  microgrids.
\end{IEEEbiography}

\begin{IEEEbiography}
[{\includegraphics[width=1.1in,height=1.2in, clip, keepaspectratio]{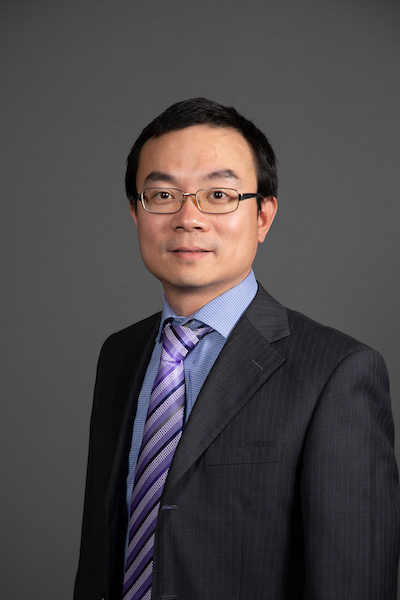}}]{Rui Bo}
(Senior Member, IEEE) received the BSEE and MSEE degrees in electric power engineering from Southeast University (China) in 2000 and 2003, respectively, and received the Ph.D. degree in electrical engineering from the University of Tennessee, Knoxville (UTK) in 2009. He is currently an Assistant Professor of the Electrical and Computer Engineering Department with the Missouri University of Science and Technology (formerly University of Missouri-Rolla). He worked as a Principal Engineer and Project Manager at Midcontinent Independent System Operator (MISO) from 2009 to 2017. His research interests include computation, optimization and economics in power system operation and planning, high performance computing, electricity market simulation, evaluation and design.
\end{IEEEbiography}

\end{document}